# Alchemical thermodynamic integration for ab initio free-energy calculations in solutions

Liangrui Wei[1], Feng Zhang[2,3], Renata M. Wentzcovitch[4-6], Kai-Ming Ho[2], Yang Sun[1*]
[1]*Department of Physics, Xiamen University, Xiamen 361005, China*
[2]*Department of Physics and Astronomy, Iowa State University, Ames, Iowa 50011, United States*
[3]*Ames National Laboratory, Ames, Iowa 50011, United States*
[4]*Department of Applied Physics and Applied Mathematics, Columbia University, New York, NY, 10027, United States*
[5]*Department of Earth and Environmental Sciences, Columbia University, New York, NY, 10027, United States*
[6]*Lamont–Doherty Earth Observatory, Columbia University, Palisades, NY, 10964, United States*

**Abstract**

We develop an alchemical thermodynamic integration scheme that couples ab initio force calculators on the fly during Monte Carlo and molecular dynamics simulations. The implementation is validated against existing hybrid-Hamiltonian approaches. The scheme yields ab initio free energies of high-pressure Fe-Ni and ambient Li-Na liquid solutions that agree with previous calculations and reproduce the experimentally observed Li-Na miscibility gap. The code has an efficiency comparable to standard ab initio molecular dynamics. These results establish this scheme as a practical alchemical-integration framework for first-principles free-energy calculations in solutions.

## 1. Introduction

The Gibbs free energy is central to computational materials science because it defines the stability and phase diagram of competing phases over a range of thermodynamic conditions [1]. At finite temperature, the Gibbs free energy is determined by both enthalpy and entropy. While ab initio molecular dynamics (AIMD) can routinely sample enthalpic contributions, entropy is not a direct observable in standard molecular dynamics (MD) or Monte Carlo (MC) simulations. As a result, obtaining free energies reliably with ab initio accuracy remains one of the main challenges in computational physics.

One common strategy is to evaluate the free energy using an explicit entropy model, in which the entropy term is defined analytically and the excess entropy is approximated from structural information, such as short-range order [2] or correlation functions [3]. While this approach has relatively low computational cost and is straightforward to combine with enthalpy calculation in AIMD, its predictive power ultimately depends on the validity of the entropy model and assumed excess term. A more rigorous approach is to use Hamiltonian thermodynamic integration (HTI), which computes free-energy *differences* by transforming a reference Hamiltonian into the target Hamiltonian. In this framework, one typically introduces reference systems with known free energy, such as analytical Lennard-Jones and Uhlenbeck-Ford systems [4,5]. However, if the reference and target systems differ dramatically, the integration path can become poorly behaved or even cross a phase transition. A remedy is to use a well-designed classical potential as an intermediate Hamiltonian to combine the analytical model with the ab initio target [6–9].

For phase-diagram construction, absolute free energies are often unnecessary. Instead, relative Gibbs free energies referenced to a suitable endmember are sufficient to determine the free energy of mixing. Therefore, an "alchemical" path that transforms a pure liquid into a liquid solution is particularly well suited for free-energy calculations of solutions, as the pure endmember provides a physically similar reference state and thus a smooth integration path [10]. Such an HTI-based strategy has been demonstrated using classical potentials with explicit forms of the interatomic interactions [10]. Extending this idea to an *ab initio* framework, however, is nontrivial. Unlike classical MD simulations, where the Hamiltonian can be directly manipulated, *ab initio* implementations require evaluating the energies and forces from density functional theory (DFT) for two systems with identical atomic configurations, which are then coupled consistently on the fly within AIMD. Thus, a practical and routine *ab initio* alchemical thermodynamic integration (ATI) scheme for free-energy calculations is needed.

In this work, we develop an ATI code for solution systems. By coupling two DFT force and energy calculators to an external MD runner, this approach enables efficient sampling along the alchemical path and direct evaluation of relative free energies, without relying on an explicit entropy ansatz or fitting an intermediate classical potential. We demonstrate the code using classical and ab initio Fe-Ni benchmarks, as well as an ab initio Li-Na liquid-solution application.

This paper is organized as follows. Section 2 describes the ATI theory, implementation, and simulation details. Section 3 presents validation and application examples, including the benchmarks with Fe-Ni and Li-Na liquid solutions. Section

*Email: yangsun@xmu.edu.cn

4 reports the computational performance. Section 5 summarizes the conclusions.

## 2. Theory and implementation

### 2.1 Alchemical thermodynamic integration

We first recap the algorithm of the ATI. Consider a solution system $A_{1-x}B_x$, where A and B are different types of atoms and $x$ is the concentration of B. The Helmholtz free energy difference between $A_{1-x}B_x$ and pure A can be calculated by introducing an auxiliary system $A_{1-x}B'_x$, in which the B' atom has the same mass as B but interacts with other atoms as A. The Helmholtz free energy difference between $A_{1-x}B'_x$ and A caused by mass difference can be written as

$$F_{\text{mass}} = F_{A_{1-x}B'_x} - F_A = k_B T\left[\frac{3}{2}x\ln\frac{m_A}{m_B} + x\ln x + (1-x)\ln(1-x)\right]. \quad (1)$$

Then, the free energy difference between $A_{1-x}B_x$ and $A_{1-x}B'_x$ can be obtained by HTI as

$$F_{\text{TI}}(V_0) = F_{A_{1-x}B_x} - F_{A_{1-x}B'_x} = \int_0^1 \left\langle \frac{\partial H_\lambda}{\partial \lambda} \right\rangle_{\lambda,NVT} \mathrm{d}\lambda = \int_0^1 \left\langle U_{A_{1-x}B_x} - U_{A_{1-x}B'_x} \right\rangle_{\lambda,NVT} \mathrm{d}\lambda, \quad (2)$$

where $\langle \ldots \rangle_{\lambda,NVT}$ denotes the MD simulation using the coupled Hamiltonian $H_\lambda = (1-\lambda)H_{A_{1-x}B'_x} + \lambda H_{A_{1-x}B_x}$ under $V_0$. In AIMD, Hamiltonians $H_{A_{1-x}B'_x}$ and $H_{A_{1-x}B_x}$ do not have an explicit analytical or numerical form. Therefore, the atomic positions can only be updated using the coupled forces as

$$f_\lambda = (1-\lambda)f_{A_{1-x}B'_x} + \lambda f_{A_{1-x}B_x}. \quad (3)$$

Thus, the ATI code evaluates forces and energies using two parallel DFT calculations on the same atomic configuration, which is the central feature of the current implementation. Combining Eqs. (1) and (2), the Helmholtz free energy difference between $A_{1-x}B_x$ and pure A under the same volume of $V_{A_{1-x}B_x}$ is given by

$$F_{A_{1-x}B_x}(V_{A_{1-x}B_x}) - F_A(V_{A_{1-x}B_x}) = F_{\text{mass}} + F_{\text{TI}}(V_{A_{1-x}B_x}). \quad (4)$$

Using the Helmholtz free energy difference, one can compute the Gibbs free energy difference. We use $V_{A_{1-x}B_x}$ and $V_A$ to represent the volume of $A_{1-x}B_x$ and $V_A$ at a given pressure $P_0$. According to $G = F + PV$ and $P = -\frac{\partial F}{\partial V}$, we obtain

$$G_{A_{1-x}B_x}(P_0) - G_A(P_0) = F_{A_{1-x}B_x}(V_{A_{1-x}B_x}) - F_A(V_{A_{1-x}B_x}) + P_0 V_{A_{1-x}B_x} - P_0 V_A - \int_{V_A}^{V_{A_{1-x}B_x}} P_A(V)\mathrm{d}V, \quad (5)$$

where $P_A(V)$ is the equation of state (EOS) of pure A. Combining the above relation, we obtain

$$G_{A_{1-x}B_x}(P_0) - G_A(P_0) = F_{\text{TI}}(V_{A_{1-x}B_x}) + F_{\text{mass}} + F_{\text{PV}}. \quad (6)$$

The $F_{\text{PV}}$ term represents the free energy contribution from the difference in EOS between $A_{1-x}B_x$ and $A$ as

$$F_{\text{PV}} = P_0 V_{A_{1-x}B_x} - P_0 V_A - \int_{V_A}^{V_{A_{1-x}B_x}} P_A(V)\mathrm{d}V. \quad (7)$$

### 2.2 Simulation details

DFT calculations were performed with the Vienna *ab initio* simulation package (VASP) [11]. The projector-augmented-wave (PAW) method was used to describe the electron-ion interaction, and the generalized gradient approximation (GGA) in the Perdew-Burke-Ernzerhof (PBE) form was employed for the exchange-correlation functional. The electronic entropy at high temperature was described by the Mermin functional [12,13]. The electronic temperature in the Mermin functional was set to match the ionic temperature. For the Fe-Ni simulations, PAW-PBE potentials were used for Fe and Ni, with $3d^6 4s^2$ valence electrons for Fe and $3d^8 4s^2$ for Ni. The plane-wave cutoff energy was set to 400 eV. Supercells containing 250 atoms were used to simulate the liquid. The time step for the AIMD and ATI simulation was 1.0 fs. The Nosé-Hoover thermostat [14] was applied with a damping time of $\tau = 0.01$ ps. For the Li-Na simulations, PAW potentials with $2s^1$ valence electrons for Li and $3s^1$ for Na were used. The plane-wave cutoff energy was set to 200 eV. Supercells containing 500 atoms were employed in the simulations. The $\boldsymbol{\Gamma}$ point was used to sample the Brillouin zone in both AIMD and ATI simulations. The time step for the AIMD and ATI simulation was 2.0 fs. The Nosé-Hoover thermostat [14] was applied with a damping time of $\tau = 0.02$ ps.

Large-scale Atomic/Molecular Massively Parallel Simulator (LAMMPS) [15] was employed to update the atomic position in the ATI code. For the coupling with Monte Carlo sampling (MCMD), we used the LAMMPS *fix atom/swap* function [15], following the atom-swap scheme described in [16]. In $Li_{0.6}Na_{0.4}$ case, a trial swap was attempted every 10 MD steps, accepted according to the Metropolis probability $P = \min[1, \exp(-\Delta E/k_B T)]$, where $\Delta E$ is the total-energy change caused by the attempted swap.

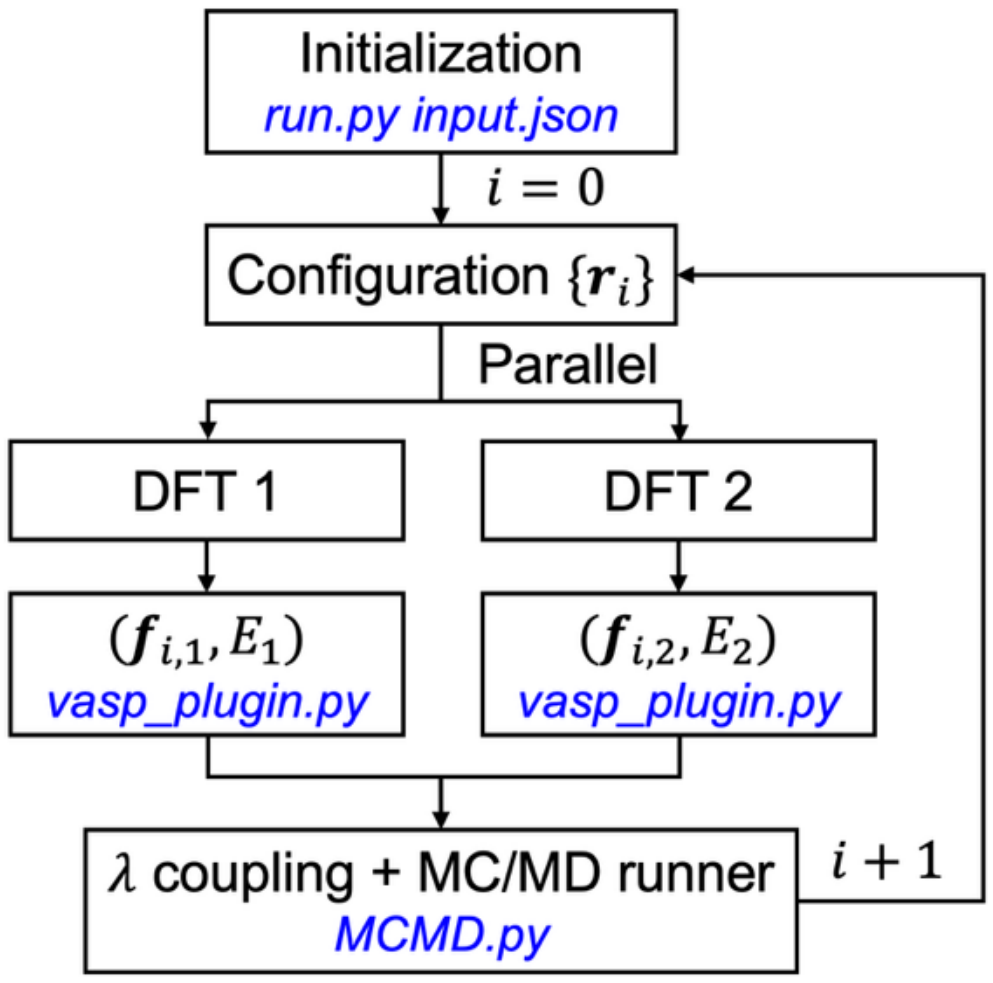

**FIG. 1** Schematic of the ATI workflow. Blue labels denote the main Python code.

### 2.3 Code framework

Figure 1 summarizes the workflow of the ATI code. The calculation is initialized from *input.json*, which defines the ATI simulation's input parameters. The initial atomic configuration, $\{\boldsymbol{r}_i\}$, is supplied to two resident DFT calculators, denoted DFT1 and DFT2. The two calculators evaluate the forces and energies for two DFT setups and return them via the backend interface. These forces and energies are passed to the MC/MD runner, where the λ-dependent coupled forces are constructed using Eq. (3) to update the atomic positions. When MC sampling is enabled, it also generates trial configurations by swapping atoms at prescribed MC intervals and uses the coupled energy to accept or reject each trial move, thereby updating the atomic configuration. This cycle continues until the user-specified number of MD steps is reached, after which the MC/MD runner and the two DFT subprocesses are terminated. In the current implementation, VASP serves as the DFT calculator, with *vasp_plugin.py* transferring structures, forces, and energies between the DFT calculator and the MC/MD runner. LAMMPS is used as the MC/MD engine, while *MCMD.py* controls MD propagation, λ coupling, and optional MC sampling. Because the two VASP processes remain active along the same ionic trajectory, wave functions and charge densities can be efficiently extrapolated from previously converged states stored in memory [17]. This avoids the overhead associated with restarting DFT calculations and improves the efficiency of electronic convergence.

## 3. Results and discussions

### 3.1 Classical Fe-Ni liquid solution

The central component of the ATI workflow is the coupling of the forces and energies from two Hamiltonians. To validate this implementation, we first perform classical MD and compare the ATI code against the existing *pair/hybrid* function in LAMMPS [15], for which an equivalent mixed-potential simulation can be performed directly for the interatomic potentials. For this benchmark, the two alchemical end states are generated from the Fe-Ni EAM potential using two LAMMPS *pair_coeff* type mappings: *Fe Fe* for the reference state and *Fe Ni* for the target state. Figure 2(a) compares the evolution of potential energies obtained from the native LAMMPS *pair/hybrid* simulation and from the ATI code. For both coupled Hamiltonians, the ATI potential-energy curves reproduce those from *pair/hybrid* simulation, demonstrating that the ATI driver correctly couples the force and energy information from the two calculators and updates atomic positions as intended.

Because *ab initio* calculations are limited in both accessible system size and trajectory length, we further used the Fe-Ni EAM benchmark to estimate the finite-size and sampling uncertainty of the TI free-energy difference. Figure 2(b) shows the calculated $\Delta F$ as a function of system size for sampling lengths from 1 to 100 ps, with the first 2 ps are discarded for equilibration. For AIMD-accessible system sizes and MD lengths (~200 atoms and10 ps), the deviation from the long-sampling and large-cell limit (10000 atoms and 200 ps) is less than 2 meV/atom.

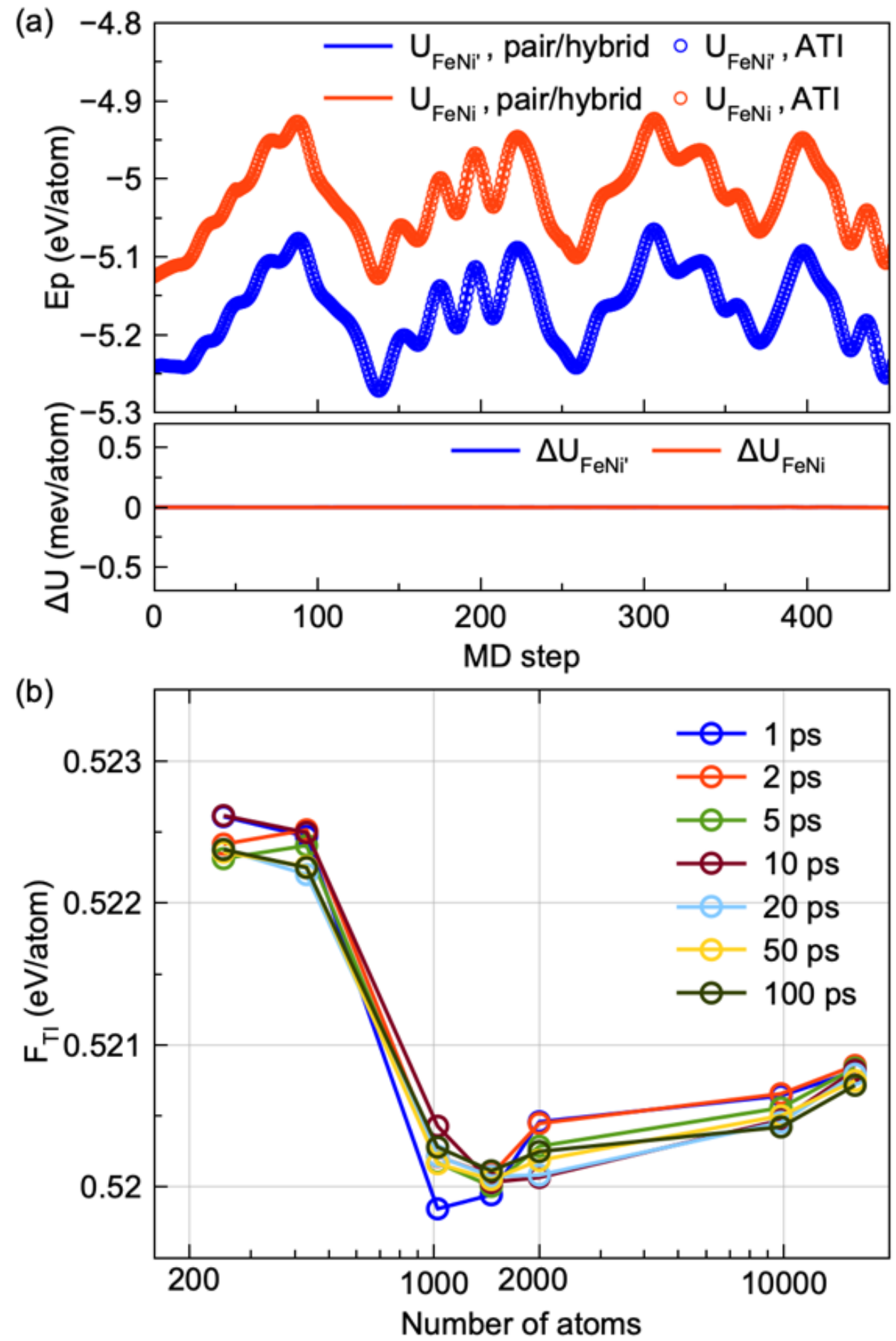

**FIG. 2** Benchmarks for the classical Fe-Ni system. (a) Comparison of the potential energies obtained from LAMMPS *pair/hybrid* function and from the current ATI code for the FeNi' and FeNi systems, with the differences shown in the lower panel, $\Delta U = U_{pair/hybrid} - U_{AATI}$. (b) Free energy from TI as a function of system size and simulation time.

### 3.2 *Ab initio* Fe-Ni liquid solution

We next compute the *ab initio* mixing free energy of $Fe_{90}Ni_{10}$ at 6000 K and 323 GPa using the ATI code. In this test, the λ-dependent force acting on each atom is obtained by coupling two DFT calculations performed with two PAW potentials, AA and AB. Here, A corresponds to Fe and B corresponds to Ni. The resulting free energy is compared with previous ab initio results obtained using the regular-solution (RS) model [18]. The alchemical transformation uses pure Fe as the reference endmember and replaces Fe by Ni′ along the alchemical path. ATI simulations are performed with six alchemical points, λ = 0, 0.2, 0.4, 0.6, 0.8, and 1.0. Each trajectory includes 0.5 ps for equilibration and 4.5 ps for ATI data collection. The MD data for λ=0.6 is shown in Fig. 3(a) as an example. The reasonable fluctuations indicate that the system reaches equilibrium and that the coupling scheme in the ATI code is stable.

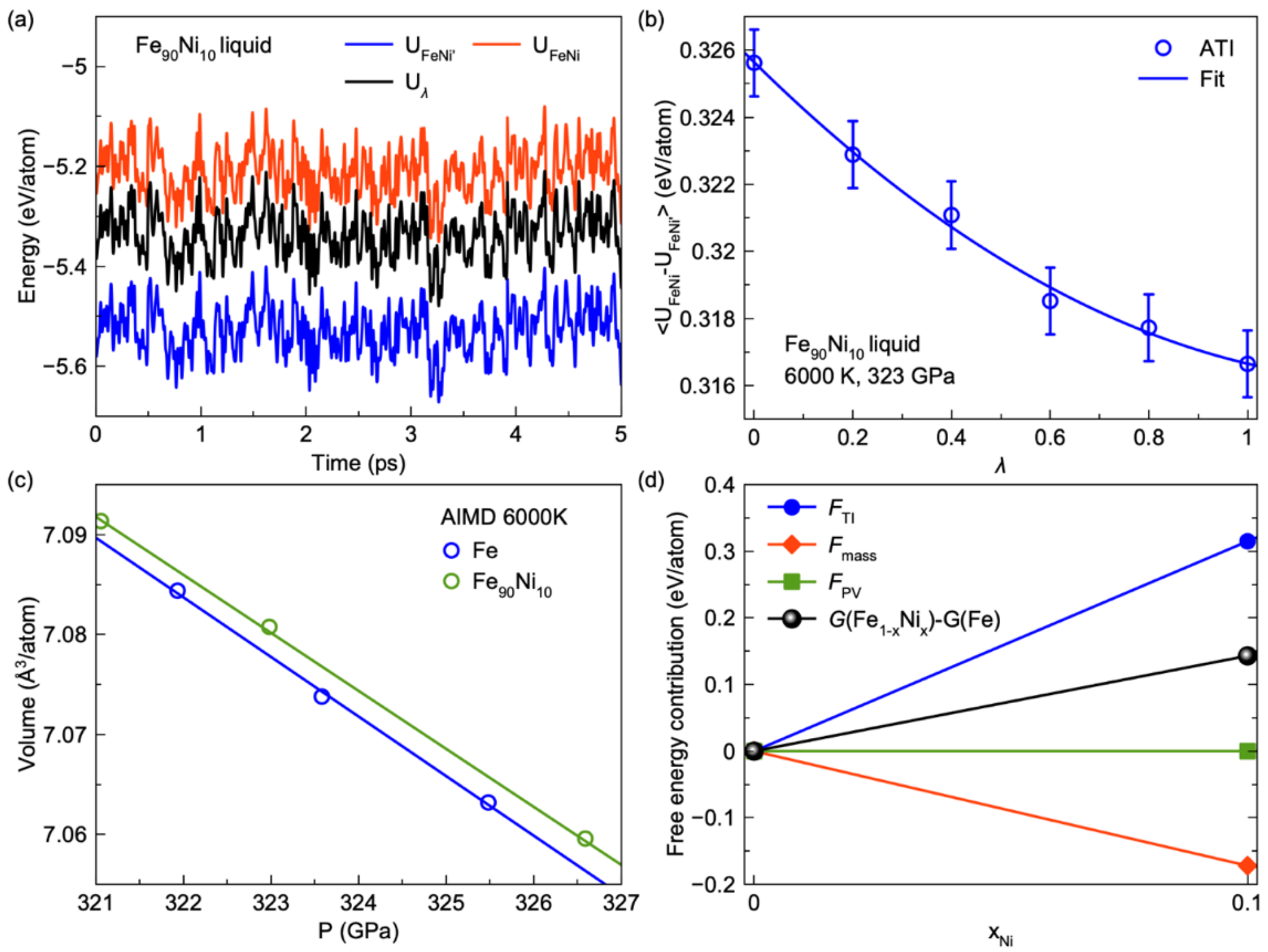

**FIG. 3**. Details of the ATI free energy calculation for $Fe_{90}Ni_{10}$. (a) Representative time evolution of the potential energies from the two Hamiltonians in one ATI simulation at 6400K and λ=0.6. (b) The integrand $\langle U_{\mathrm{FeNi}} - U_{\mathrm{FeNi'}}\rangle$ as a function of $\lambda$ for liquid $Fe_{1-x}Ni_x$ at 6400 K. The solid line is a quadratic fitting curve. (c) P-V relation of the Fe and $Fe_{90}Ni_{10}$ liquids at 6400 K. Each pressure was computed by 5.0 ps of AIMD. (d) Contributions from $F_{\mathrm{TI}}$, $F_{\mathrm{mass}}$, and $F_{\mathrm{PV}}$ to the Gibbs free energy difference between Fe and $Fe_{90}Ni_{10}$ liquids.

Figure 3(b) shows the alchemical integrand $\langle U_{\mathrm{FeNi}} - U_{\mathrm{FeNi'}}\rangle$ as a function of λ. The error bar of 0.001 eV/atom is estimated from convergence tests of the alchemical integrand at fixed λ. The $F_{\mathrm{TI}}$ term is obtained by integrating the fitted second-order polynomial curve using Eq. (2), yielding $F_{\mathrm{TI}}$= 0.315 ± 0.001 eV/atom. The $F_{\mathrm{PV}}$ term is calculated from Eq. (7) and the P–V relation shown in Fig. 3(c). $F_{\mathrm{PV}}$ contributes only 0.00004 eV/atom in this system, mainly because Fe and Ni have very similar atomic volumes under these conditions. The total free-energy difference is therefore dominated by $F_{\mathrm{TI}}$ and $F_{mass}$. $F_{mass}$ is −0.172 eV/atom according to Eq. (1). Combining these terms using Eq. (6), we obtain a Gibbs free energy difference of 0.143 ± 0.001 eV/atom for $Fe_{90}Ni_{10}$ relative to pure Fe. In Ref. [18], the regular-solution (RS) model, based on AIMD simulations of the same system size, also yields a corresponding value of 0.143 eV/atom at 6000 K and 323 GPa. The excellent agreement between the two methods confirms the accuracy of the present ATI scheme. It also suggests that liquid $Fe_{90}Ni_{10}$ can be well described by the RS model.

### 3.3 *Ab initio* Li-Na liquid solution

We next apply ATI to compute the free energy of the Li–Na liquid solution at ambient pressure, for which previous ab initio calculations and experimental data are available for comparison [19–21]. The experimental Li–Na phase diagram exhibits a pronounced low-temperature miscibility gap [19,20]. Huang *et al.* computed the free energy of Li–Na phase diagram by combining enthalpies from AIMD simulations with an entropy model [21]. We first compute the $F_{\mathrm{PV}}$ term using Eq. (7), which requires the equilibrium volume of $Li_xNa_{1-x}$ at 0 GPa and 473 K, as well as the P(V) curve of pure Na. AIMD simulations are performed for $Li_xNa_{1-x}$ liquids with $x_{\mathrm{Li}} = 0.2$, 0.4, 0.6, 0.8, and 1.0 at five different volumes. The resulting P–V data are fitted using the Birch–Murnaghan equation of state, which yields the equilibrium volume at 0 GPa for each composition. These equilibrium volumes are shown as a function of Li concentration in Fig. 4(b). Unlike Fe–Ni, the volume changes significantly with Li content because Li and Na have substantially different atomic volumes. The present volumes are consistent with previous AIMD calculations [21], but are systematically lower than the experimental values by about 2% [19]. This discrepancy can be attributed to exchange-correlation effects in DFT, which are known to underestimate equilibrium volumes [22,23]. Because the equilibrium volumes of liquid $Li_xNa_{1-x}$ span a wide range, the P(V) curve of pure Na must also cover this range so that the integration in Eq. (7) can be evaluated. The corresponding AIMD results are shown in Fig. 4(c) and are well fitted by the third-order Birch–Murnaghan equation of state,

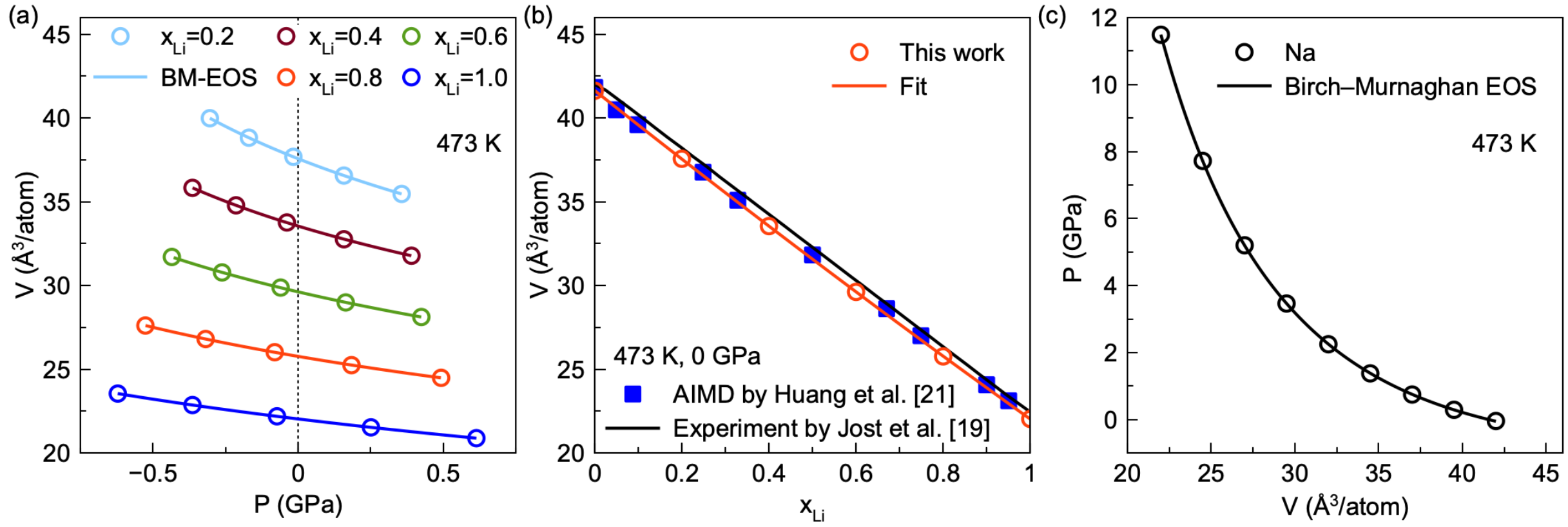

**FIG. 4** P–V relations of liquid $Li_xNa_{1-x}$ at 473 K. (a) P–V relations near 0 GPa for different Li concentrations. Each data point was obtained from 1 ps of equilibration followed by 8 ps of AIMD simulation. The lines are fitted with the Birch–Murnaghan equation of state. (b) Volume as a function of Li concentration, compared with the experimental data of Jost et al. [19] and the AIMD results of Huang et al. [21]. The red line is a quadratic fit to get volume at any composition. (c) P–V relation of liquid Na covering the volume range of liquid $Li_xNa_{1-x}$. The line is the third-order Birch–Murnaghan EOS.

$$P(V) = \frac{3B_0}{2}\left[\left(\frac{V_0}{V}\right)^{7/3} - \left(\frac{V_0}{V}\right)^{5/3}\right]\left\{1 + \frac{3}{4}(B_0' - 4)\left[\left(\frac{V_0}{V}\right)^{2/3} - 1\right]\right\}, \quad (8)$$

where the fitted parameters are $B_0$=5.143 GPa, $B_0' = 3.943$, and $V_0$=41.552 Å$^3$/atom. This expression for $P_{Na}(V)$ is then used to compute $F_{\rm PV}$ term in Eq. (7).

Based on the composition-dependent equilibrium volumes, we performed ATI simulations for $x_{\rm Li}$ = 0.1, 0.15, 0.2, 0.4, 0.6, 0.8, 0.85, 0.9, 0.95, and 1.0, with $\lambda$ = 0, 0.25, 0.5, 0.75, and 1.0. Pure Na was used as the reference endmember, and Na atoms were gradually replaced by Li along the alchemical path. Each trajectory consisted of 1 *ps* of equilibration followed by 10 *ps* of ATI sampling. Figure 5(a) shows the alchemical integrand, $\langle U_{\rm LiNa} - U_{\rm LiNa'}\rangle$, for different Li concentrations. All integrands vary smoothly with λ, resulting in well-converged integration. For $Li_{0.6}Na_{0.4}$, where the liquid tends to phase separate, we used the MCMD runner in the ATI calculations to test convergence. Two independent MCMD runs are performed for 10 ps and yielded nearly identical energies, as shown in Fig. 5(b). For $\lambda$ values from 0 to 0.75, the MD and MCMD results are nearly indistinguishable. However, at $\lambda$ = 1.0, corresponding to the real $Li_{0.6}Na_{0.4}$ system, the MCMD simulations yield an energy that is 0.004 eV/atom lower than that obtained from MD. This difference can be attributed to the ability of MCMD to drive the system closer to equilibrium than MD, particularly when the system tends to phase separate. A similar energy difference between MD and MCMD was also reported in Ref. [21]. Because this system tends to phase separate, it is important to suppress this separation during the calculation to obtain the free energy of the homogeneous mixed phase. In the present simulations, the system remains in a spatially uniform single phase. This is confirmed by the pair correlation functions (PCFs), which show only minor changes from the Li′Na system (pure Na interactions) to the real Li–Na system, as shown in the inset of Fig. 5(b).

Figure 5(c) shows the contributions of the $F_{\rm TI}$, $F_{\rm mass}$, and $F_{\rm PV}$ terms to the Gibbs free-energy difference, $\Delta G(x) = G(\mathrm{Li}_x\mathrm{Na}_{1-x}) - G(\mathrm{Na})$, between $Li_xNa_{1-x}$ and pure Na. The $F_{\rm TI}$ contribution is negative and increases in magnitude with increasing Li content, whereas the $F_{\rm PV}$ term is positive and compensates for much of this decrease. The $F_{\rm PV}$ contribution is much larger than that in the Fe–Ni system. Using $\Delta G(x)$, the Gibbs free energy of mixing can be calculated as $G_{\rm mix}(x) = \Delta G(x) - x\Delta G(1)$, as shown in Fig. 5(d). The ATI results closely reproduce the 473 K free energies reported by Huang et al. [21] at both high and low Li concentrations, with deviations of about 1 meV/atom. The MCMD-ATI free energy at $x_{\rm Li}$=0.6 is about 0.5 meV/atom lower than the corresponding MD-ATI value. The free-energy curve can be fitted well using the Redlich–Kister (RK) expression [24]. A large portion of the free-energy curve lies above the convex hull over the composition range $x_{\rm Li}$ of 0.1–0.95, indicating thermodynamic instability with respect to liquid–liquid phase separation. The resulting phase-separation interval is consistent with experimental observations, which show phase separation between $x_{\rm Li}$ at 0.15 and 0.95 [19,20]. The agreement is excellent on the Li-rich side, whereas the Na-rich boundary shows a slightly larger deviation from experiment. This discrepancy can be attributed to the DFT accuracy limit of about 1 meV/atom and the finite-size error of approximately 2 meV/atom shown in Fig. 2(b), since such energy uncertainties can readily shift the calculated phase boundary to match experiment.

## 4. Computational performance

We evaluate the computational performance of the ATI implementation. Figure 6 compares the wall time per MD step as a function of system size for AIMD, ATI, and tests with classical potentials. We use wall time rather than total

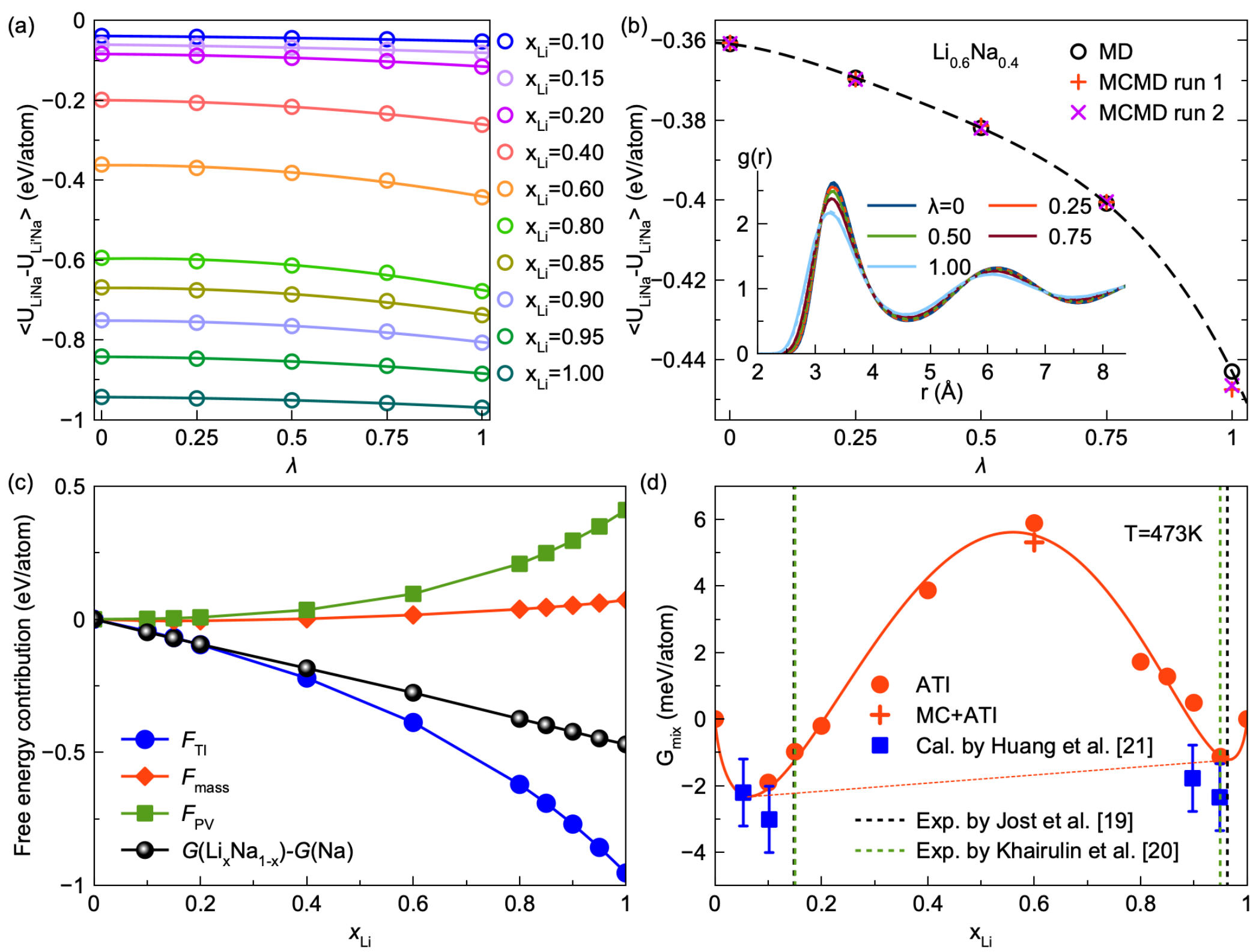

**FIG. 5** Free-energy calculations for Li–Na liquids at 473 K. (a) Alchemical integrand, $\langle U_{\mathrm{LiNa}} - U_{\mathrm{Li'Na}}\rangle$, as a function of λ for different Li concentrations. The solid lines are quadratic fits. (b) ATI calculations at $x_{\mathrm{Li}}$=0.6 using MD and MCMD simulations. Each run lasts 10 ps. The line is a quadratic fit to the ATI-MD data. The inset shows the pair correlation functions obtained from MCMD (solid lines) and MD (dashed lines) simulations, which overlap for in λ values. (c) Contributions of $F_{\mathrm{TI}}$, $F_{\mathrm{mass}}$, and $F_{\mathrm{PV}}$ to the Gibbs free-energy difference relative to pure Na as a function of Li concentration. (d) Gibbs free energy of mixing, $G_{\mathrm{mix}}(x)$. The solid line is a fit using the Redlich–Kister (RK) model, $G_{mix}(x) = x(1-x) * (2.9776 + 0.6524 * x) + 473 * k_B * (xln(x) + (1-x)ln(1-x))$ in eV/atom. The horizontal dotted line indicates the phase-separation boundary predicted by the RK model. The blue symbols are data from the previous calculations of Huang et al. [21]. The vertical dotted lines indicate the experimentally measured phase-separation boundaries [19,20].

CPU time because ATI runs two DFT calculators in parallel; therefore, its CPU usage is twice that of a standard AIMD simulation, whereas the wall time reflects the practical time-to-solution.

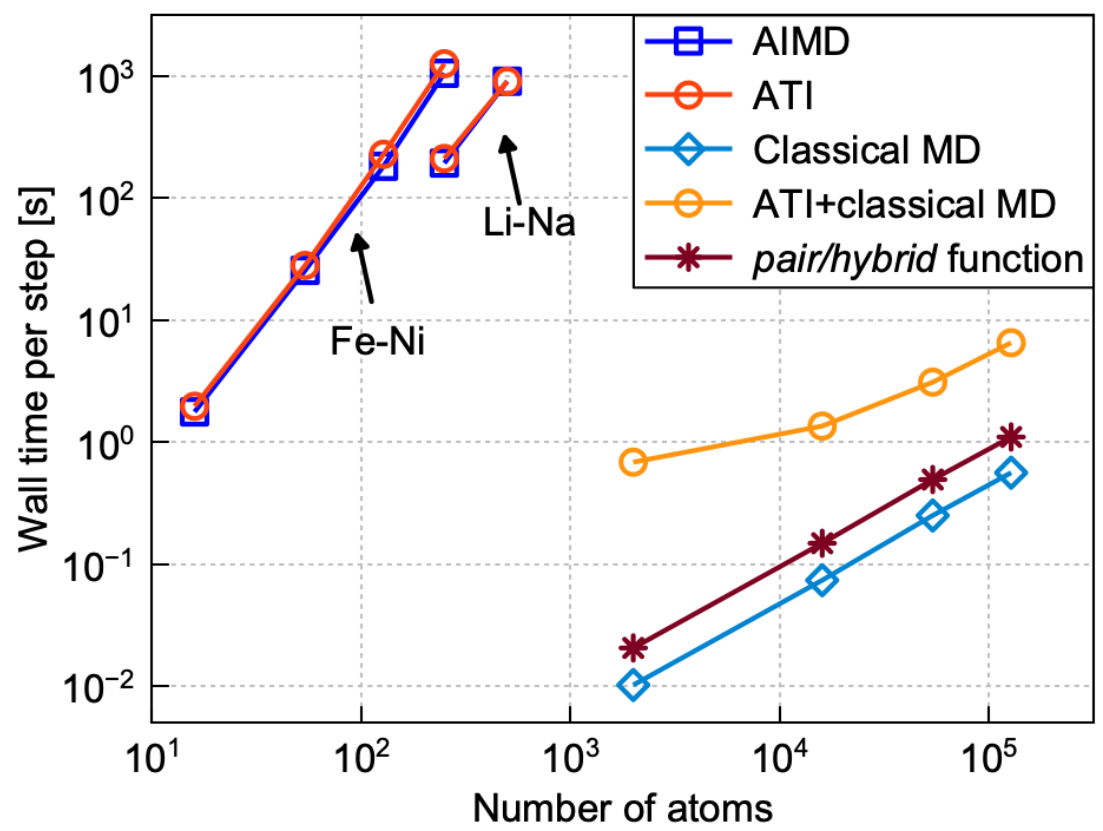

**FIG. 6** Computational efficiency of the ATI implementation. Wall time per MD step is shown as a function of system size for different implementations.

For the present implementation, the main overhead arises from passing energies, forces, and atomic configurations through the external *vasp_plugin.py* interface, which involves callback operations and data exchange between the force calculators and the MD runner. For ab initio simulations, this overhead is negligible compared with the cost of the electronic structure calculations. As a result, the wall time of ATI remains close to that of standard AIMD over the tested system sizes. In contrast, the overhead is more apparent with classical force fields, where force evaluation is inexpensive. The ATI coupled to classical MD is therefore about one order of magnitude slower than the native LAMMPS pair/hybrid calculation, which directly combines explicit potential forms inside LAMMPS. Since ATI is designed primarily for ab initio Hamiltonians that cannot be directly coupled within a classical MD engine, its near-AIMD wall-time performance is satisfactory for first-principles applications.

## 5. Conclusions

We have developed an ATI code for evaluating free energies in solution systems. The implementation couples two resident force and energy calculators to an external MC/MD runner, allowing the forces and energies from two alchemical end states to be combined on the fly during MC/MD simulations. We validated the implementation using both classical and ab initio benchmarks. For the classical Fe-Ni liquid, the ATI results reproduce those from the native LAMMPS pair/hybrid implementation, confirming that the code correctly couples forces and energies from two Hamiltonians. For Fe-Ni liquid under high-pressure and high-temperature conditions, ATI yields a Gibbs free energy in excellent agreement with previous ab initio results based on the regular-solution model. We also applied the method to the Li-Na liquid solution, where the large volume difference between Li and Na makes the EOS-related contribution important. The calculated free energies are consistent with previous ab initio data and well capture the experimentally observed tendency toward liquid-liquid phase separation. Performance tests show that the wall time of ATI remains close to that of standard AIMD.

Because the method only requires calculator-returned energies and forces, it is not limited to Hamiltonians with explicit analytical or numerical forms. This feature makes ATI applicable to general first-principles and machine-learning force calculators, where direct manipulation of the Hamiltonian is otherwise difficult. Looking forward, the same calculator-based design can be extended to other DFT packages and to machine-learning potentials, providing a general alchemical-integration framework for free-energy calculations with MC and MD sampling.

**Acknowledgements**
Work at Xiamen University was supported by the National Natural Science Foundation of China (Grants Nos. 42550120 and T2422016). Work at Ames National Laboratory was supported by the U.S. Department of Energy (DOE), Office of Science, Basic Energy Sciences, Materials Science and Engineering Division through the Computational Materials Science Center program. Ames National Laboratory is operated for the U.S. DOE by Iowa State University under contract No. DE-AC02-07CH11358. R.M.W acknowledges support from the Gordon & Betty Moore Foundation (doi.org/10.37807/GBMF12801)